\documentclass[%
 aip, jap,
 amsmath,amssymb,
reprint,%
]{revtex4-1}

\usepackage{graphicx}
\usepackage{dcolumn}
\usepackage{bm}

\usepackage[utf8]{inputenc}
\usepackage[T1]{fontenc}
\usepackage{amsthm}
\usepackage{color}
\usepackage{upgreek}
\usepackage{comment}
\usepackage{ulem}

\newtheoremstyle{query}%
{}{}
{\color{red}}
{}
{\sffamily\bfseries}{:}{12pt}
{}
\theoremstyle{query}

\newcommand{\red}[1]{{\textcolor{black}{#1}}}

\begin{document}

\preprint{AIP/123-QED}

\title[]{Epitaxial growth and orientation-dependent anomalous Hall effect of noncollinear antiferromagnetic Mn$_3$Ni$_{0.35}$Cu$_{0.65}$N films}

\author{R. Miki}
\affiliation{Department of Materials Physics, Nagoya University, Nagoya 464-8603, Japan}
\author{K. Zhao}
\affiliation{Experimentalphysik VI, Center for Electronic Correlations and Magnetism, University of Augsburg, 86159 Augsburg, Germany}
\author{T. Hajiri}
 \email{t.hajiri@nagoya-u.jp}
\affiliation{Department of Materials Physics, Nagoya University, Nagoya 464-8603, Japan}
%
%
\author{P. Gegenwart}
\affiliation{Experimentalphysik VI, Center for Electronic Correlations and Magnetism, University of Augsburg, 86159 Augsburg, Germany}
\author{H. Asano}
\affiliation{Department of Materials Physics, Nagoya University, Nagoya 464-8603, Japan}
%

\date{\today}

\begin{abstract}
We report the growth of noncollinear antiferromagnetic (AFM) Mn$_3$Ni$_{0.35}$Cu$_{0.65}$N films and the orientation-dependent anomalous Hall effect (AHE) of (001) and (111) films due to nonzero Berry curvature.
We found that post-annealing at 500\,$^\circ$C can significantly improve the AHE signals, though using the appropriate post-annealing conditions is important.
\red{The AHE and magnetization loops  show sharp flipping at the coercive field in (111) films, while (001) films are hard to saturate by a magnetic field.}
The anomalous Hall conductivity of (111) films is an order of magnitude larger than that of (001) films.
The present results provide not only a better understanding of the AHE in Mn$_3X$N systems but also further opportunities to study the unique phenomena related to noncollinear AFM.
\end{abstract}

\maketitle
\section{INTRODUCTION}
Antiferromagnetic (AFM) spintronics is an interesting aspect of spintronics because extraordinary properties are expected, such as the absence of a stray field, terahertz spin dynamics, low electrical current switching, and robustness against external perturbations.\cite{AFM_spintronics1, AFM_spintronics2, AFM_spintronics3, Romain_Nature}
Currently, electrical current switching as well as detecting of the AFM N\'eel vector has been realized.\cite{CuMnAs_SOT, CuMnAs_SOT2, Mn2Au_SOT, Mn2Au_SOT2, PRL_NiO_SOT, Lorenzo_NiO_SOT, Moriyama_NiO_SOT}
In these studies of collinear AFMs, the electrical signal of the N\'eel vector is, however, small via anisotropic magnetoresistance\cite{Mn2Au_SOT} (a few percent) or spin Hall magnetoresistance (${\sim}10^{-2}$\%).\cite{SMR1} 
The anomalous Hall effect (AHE) is promising for detecting a N\'eel vector with a larger electrical signal. 
The AHE is proportional to the magnetization in conventional ferromagnets.\cite{AHE1} 
In a noncollinear AFM, symmetry is broken due to geometric frustration; however, although the magnetization is quite small, of the order of $10^{-3}\mu_{\rm B}$ per atom, the AHE can be obtained due to nonzero Berry curvature.\cite{Hua_Chen_PRL} 
In practice, a large AHE is observed in hexagonal Mn$_3$Sn/Ge, where the presence of Weyl points near the Fermi energy increases the Berry curvature and thus, enhances the AHE.\cite{Mn3Sn, Mn3Ge}
On the other hand, the cubic Mn$_3$Ir/Pt is theoretically predicted to exhibit the finite AHE,\cite{Hua_Chen_PRL} and actually, the AHE is observed in Mn$_3$Pt.\cite{Mn3Pt} 
In these materials, Mn atoms commonly form a kagome lattice in the (0001) and (111) planes, respectively. 
Towards the application in AFM spintronics, thin films of these materials have been actively grown.\cite{Mn3Sn_poly1, Mn3Sn_poly2, Mn3Sn_poly3, Mn3Sn_epi1, Mn3Sn_epi2, Mn3Sn_epi3, Mn3Ir_epi}

The antiperovskite nitride Mn$_3X$N system is a noncollinear AFM with a kagome lattice in the (111) plane. 
Its AFM spin order ($\Gamma_{4g}$ and $\Gamma_{5g}$) varies depending on the composition of the $X$ site. 
According to symmetry analysis, when the $\Gamma_{4g}$ order or the $\Gamma_{5g}$ order with strain is realized, nonzero Berry curvature in the momentum space is expected, which may result in the finite AHE\cite{Mn3NiN_AHE} and giant anomalous Nernst effect.\cite{Nernst}
On the other hand, in Mn$_3X$N systems, the low electrical current switching of AFM moments has been demonstrated in Mn$_3$GaN/ferromagnets\cite{MGN_STT} and Mn$_3$GaN/Pt bilayers.\cite{MGN_SOT}
In addition, Mn$_3X$N has unusual physical properties, such as giant negative thermal expansion,\cite{NTE1, NTE2} temperature-independent resistivity,\cite{zeroR} and barocaloric effects.\cite{barocaloric1, barocaloric2}
Therefore, Mn$_3X$N systems are interesting materials for noncollinear AFM phenomena as well as AFM spintronics.
So far, there have been no reports regarding the growth of single crystals of Mn$_3X$N, though epitaxial Mn$_3X$N films have been produced.\cite{MGN_STT, MGN_SOT, MGN_growth, Mn3XN_growth1}. 
In order to fully understand and exploit these phenomena, epitaxial Mn$_3X$N films with nonzero Berry curvature are desired.

Our previous work demonstrated the AHE in Mn$_3$Ni$_{1-x}$Cu$_x$N systems due to nonzero Berry curvature.\cite{Cu-MNN_AHE}
In Mn$_3$Ni$_{1-x}$Cu$_x$N polycrystalline powders, although there is no clear magnetization hysteresis loop in Mn$_3$NiN, we found that Cu doping leads to clear hysteresis, even in an AFM region, due to the canted $\Gamma_{4g}$ order. 
In the same way, Mn$_3$NiN~(111) films show no AHE and Mn$_3$Ni$_{0.35}$Cu$_{0.65}$N~(111) films have a relatively large AHE of ${\sim}21.5$~($\Omega$~cm)$^{-1}$ with a small canted magnetization of the order of $10^{-3}\mu_{\rm B}$ per Mn atom. 
These results indicate that the Cu doing is essential in stabilizing the $\Gamma_{4g}$ order and therefore the AHE is realized in Mn$_3$Ni$_{0.35}$Cu$_{0.65}$N.
On the other hand, while the film-orientation dependence of AHE has been theoretically predicted,\cite{MGN_AHE} we have studied only Mn$_3$Ni$_{0.35}$Cu$_{0.65}$N~(111) films, and moreover, the details of thin film growth of Mn$_3$Ni$_{1-x}$Cu$_x$N have not yet been reported.

In this paper, we report the growth of thin films, and describe the magnetic and transport properties of Mn$_3$Ni$_{0.35}$Cu$_{0.65}$N~(001) and (111) films. 
With optimized growth conditions, the quality of the Mn$_3$Ni$_{0.35}$Cu$_{0.65}$N thin films is \red{still} poor and almost no AHE is observed. 
Post-annealing significantly improved the film quality, and the dependence of the AHE on the post-annealing conditions is clear. 
In post-annealing conditions optimized to produce the AHE, a perpendicular magnetization hysteresis loop due to canting AFM moments is obtained in (111) films with 0.003$\mu_{\rm B}$ per Mn atom at 50~K, while (001) films \red{is hard to saturate by a magnetic field}. 
In the same way, AHE hysteresis loops \red{show a sharp sign change around a coercive field} in (111) films below the N\'eel temperature, while those of (001) films \red{show a gradual sign change}. 
We found that the anomalous Hall conductivity (AHC) of (111) films is an order of magnitude larger than that of (001) films.

\section{EXPERIMENTAL DETAILS}
Epitaxial Mn$_3$Ni$_{0.35}$Cu$_{0.65}$N films (hereinafter referred to as Cu-MNN films) were prepared on MgO (001) and (111) substrates by reactive magnetron sputtering using a Mn$_{3}$Ni$_{0.35}$Cu$_{0.65}$ target under an Ar/N$_2$ atmosphere.
The nitrogen gas ratio (N$_2\%$) was controlled by two independent mass-flow controllers of pure Ar gas and an Ar~+~10$\%$~N$_2$ gas mixture, respectively.
The total pressure during film growth was 2.0~Pa.
The film growth rate was fixed to be 1.3~nm/min.
The crystal structure was analyzed using out-of-plane and in-plane X-ray diffraction (XRD) measurements with Cu~$K\alpha$ radiation.
The magnetic properties were characterized by superconducting quantum interference device magnetometry using a MPMS device (Quantum Design). 
Electronic transport data were obtained by a conventional four-probe method using a  PPMS device (Quantum Design).

\section{RESULTS AND DISCUSSION}
\begin{figure}[t]
\begin{center}
\includegraphics[width=\linewidth]{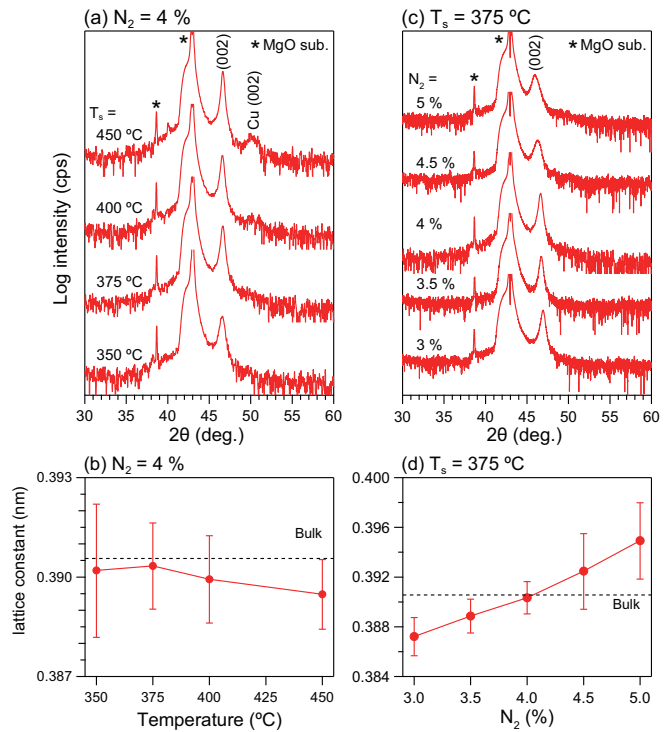}
\caption{Out-of-plane XRD results for Cu-MNN films on MgO (001) substrates.
(a) XRD profiles vs.\ $T_s$ around (002) peaks under N$_2=4.0\%$. 
(b) Lattice constant $c$ vs.\ substrate temperature.
The dashed line indicates the bulk lattice constant.\cite{Cu-MNN_AHE}
(c) XRD profiles vs.\ N$_2\%$ around (002) peaks at $T_s=375\,^{\circ}$C.
(d) Lattice constant $c$ vs.\ N$_2\%$.
}
\label{fig:one}
\end{center}
\end{figure}

Figure~\ref{fig:one}(a) shows the dependency of out-of-plane XRD patterns on substrate temperature ($T_s$) around (002) peaks of Cu-MNN films on MgO (001) substrates under an N$_2=4.0\%$ atmosphere.
Only the Cu-MNN (002) peaks have Bragg peaks below $T_s=375\,^{\circ}$C, while additional Cu (002) peaks appear above 400\,$^\circ$C.
The out-of-plane lattice constant $c$ is shown in Fig.~\ref{fig:one}(b).
The lattice constant approaches the bulk value (0.3906~nm, Ref.~\onlinecite{Cu-MNN_AHE}) at $T_s=375\,^{\circ}$C and becomes shorter for higher $T_s$.
The dependency of out-of-plane XRD profiles on N$_2\%$ at $T_s=375\,^{\circ}$C is shown in Fig.~\ref{fig:one}(c).
Only the (002) Cu-MNN peaks have Bragg peaks in the N$_2\%$ region shown.
As shown in Fig.~\ref{fig:one}(d), the lattice constant becomes longer for higher N$_2\%$.
In the same way, we studied the growth of Cu-MNN films for various combinations of $T_s$ and N$_2\%$, and concluded that $T_s=375\,^{\circ}$C and N$_2=4.0\%$ are best for growing Cu-MNN films.

\begin{figure*}
\begin{center}
\includegraphics[width=0.8\linewidth]{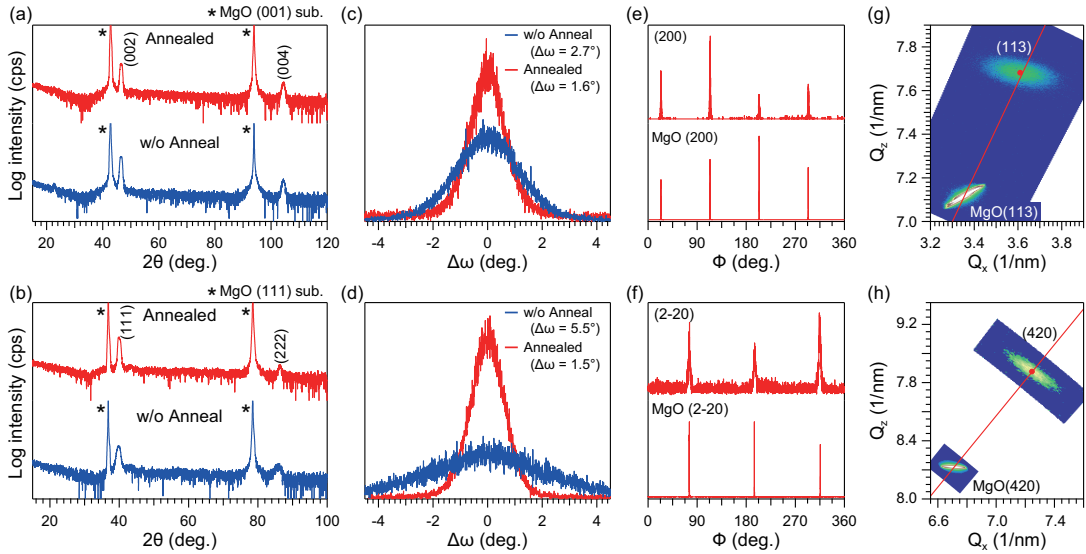}
\caption{
XRD results of Cu-MNN films on MgO (001) (upper panels) and (111) (lower panels) substrates.
Comparison of out-of-plane XRD profiles of as-grown and gas annealed Cu-MNN (a) (001) and (b) (111) films.
Rocking curves of (c) (002) and (d) (111) reflections.
In-plane $\phi$ scans of gas annealed Cu-MNN (e) (001) and (f) (111) films.
Reciprocal space maps around (113) and (420) reflections of (g) (001) and (h) (111) films. 
The solid lines and filled circles indicate the relaxation lines and film peak positions, respectively.
The gas annealed data for the (111) films are adapted from Ref.~\onlinecite{Cu-MNN_AHE}.
}
\label{fig:two}
\end{center}
\end{figure*}

Out-of-plane XRD profiles of Cu-MNN films on MgO (001) and (111) substrates for the full angle range are shown in Figs.~\ref{fig:two}(a) and~\ref{fig:two}(b), respectively.
As-grown Cu-MNN films show only the (002) and (111) peak series without any impurity peaks on MgO (001) and (111), respectively, which indicates that (001)- and (111)-oriented Cu-MNN grows on MgO (100) and (111) substrates, respectively.
As shown in Figs.~\ref{fig:two}(c) and~\ref{fig:two}(d), the full width half maxima (FWHM) of rocking curves of (002) and (111) reflections are 2.7$^{\circ}$ for (001) film and 5.5$^{\circ}$ for (111) films, implying that the film is low quality.
In situ post-annealing was used to improve the film quality.
After being grown, the Cu-MNN films were heated to 500$\,^{\circ}$C at a rate of 200$\,^{\circ}$C per hour, and kept for 30~min under the same atmosphere used for film growth (hereinafter referred to as gas anneal).
After the gas anneal, no significant change was observed in the out-of-plane XRD profiles, as shown in Figs.~\ref{fig:two}(a) and~\ref{fig:two}(b).
On the other hand, as shown in Figs.~\ref{fig:two}(c) and~\ref{fig:two}(d), the FWHMs of rocking curves of (002) and (111) reflections are much narrower due to the gas annealing, being 1.6$^{\circ}$ and 1.5$^{\circ}$, respectively.
The post-annealing and annealing conditions affect the transport properties, which is discussed later.

\begin{figure}[b]
\begin{center}
\includegraphics[width=0.95\linewidth]{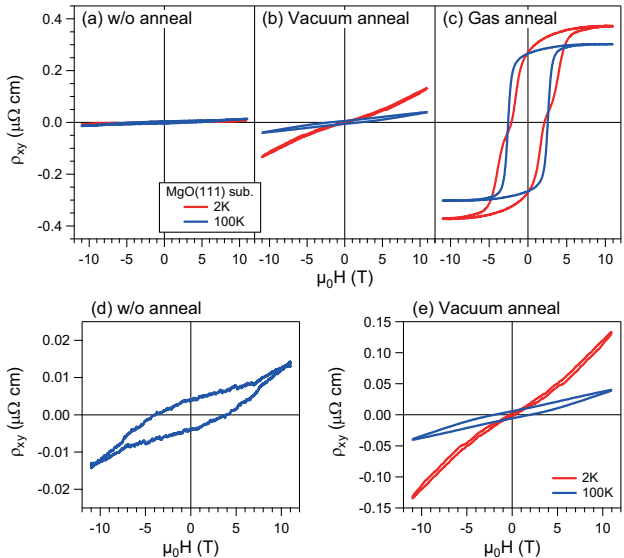}
\caption{
Dependence of the Hall resistivity of (111) films on the annealing conditions: (a) without annealing, with annealing (b) under a vacuum (vacuum anneal), and (c) under the same atmosphere used for growing the film (gas anneal).
\red{The enlarged Hall resistivity of films without and with annealing is shown in panel (d) and (e).} 
The gas annealing data for the (111) films are adapted from Ref.~\onlinecite{Cu-MNN_AHE}.
}
\label{fig:three}
\end{center}
\end{figure}

Figures~\ref{fig:two}(e) and~\ref{fig:two}(f) show in-plane $\phi$ scans of gas annealed Cu-MNN (001) and (111) films, respectively\red{, where (2-20) peaks of (111) films are acquired at $\chi=20.0^{\circ}$ to distinguish three-fold single or six-fold twin domains}.
There are clear four- and three-fold symmetries with respect to MgO peaks, indicating that the (001) and (111) films were epitaxially grown \red{with single domain} on MgO (001) and (111) substrates, respectively. 
Figures~\ref{fig:two}(g) and~\ref{fig:two}(h) are reciprocal space maps of Cu-MNN (001) and (111) films around (113) and (420) reflections. 
The (113) and (420) peaks are observed almost on the relaxation lines, indicating that the Cu-MNN films have relaxed to the bulk lattice constant.
These results indicate the successful growth of epitaxial Cu-MNN (001) and (111) films.

The Hall resistivity $\rho_{xy}$ as a function of the magnetic field of Cu-MNN (111) films for several annealing conditions are shown in Fig.~\ref{fig:three}.
As shown in Figs.~\ref{fig:three}(a) \red{and \ref{fig:three}(d)}, quite small Hall hysteresis loops are observed in films that have not been annealed.
After annealing under a vacuum (vacuum anneal) \red{as shown in Figs.~\ref{fig:three}(b) and \ref{fig:three}(e)}, the Hall signals become bigger, though the Hall hysteresis loops are still unclear.
In contrast, after annealing under the same atmosphere as used for film growth (gas anneal) \red{presented in Fig.~\ref{fig:three}(c)}, the Hall signals show clear hysteresis loops.
Although the XRD results for the lattice constants of films annealed under a vacuum or gas are almost the same (0.39020 and 0.39012~nm, respectively), these results may suggest the importance of the N concentration or hybridization between Mn and N atoms\cite{hybridization} in obtaining large AHE signals.

\begin{figure}[t]
\begin{center}
\includegraphics[width=\linewidth]{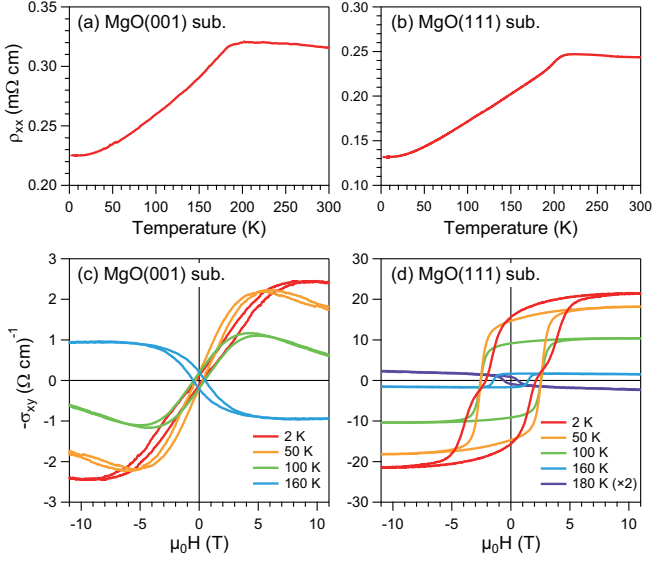}
\caption{
Temperature-dependent resistivity of (a) (001) and (b) (111) films.
AHC loops of (c) (001) and (d) (111) films at various temperatures.
The gas annealing data for the (111) films are adapted from Ref.~\onlinecite{Cu-MNN_AHE}.
}
\label{fig:four}
\end{center}
\end{figure}

Figures~\ref{fig:four}(a) and~\ref{fig:four}(b) show the temperature-dependent resistivity $\rho_{xx}$ of gas-annealed Cu-MNN (001) and (111) films, respectively.
There are clear kinks around 200~K.
This is typical behavior of an AFM transition in antiperovskite nitrides Mn$_3X$N.\cite{Neel_MGN1} 
The estimated transition temperature is close to the bulk N\'eel temperature.\cite{Cu-MNN_AHE}
The AHC $\sigma_{xy}$ as a function of the magnetic field of Cu-MNN (001) and (111) films after gas annealing is shown in Figs.~\ref{fig:four}(c) and~\ref{fig:four}(d), respectively.
Here, the AHC is calculated using $\sigma_{xy}=-\rho_{xy}/\rho^{2}_{xx}$.
\red{AHC loops are obtained in both Cu-MNN (001) and (111) films, but the different AHC loop shapes are observed; a sharp sign change in (111) films and a gradual sign change in (001) films.}
The sign of the Hall signal depends on the temperature for  both (001) and (111) Cu-MNN films. 
Those transition temperatures are slightly different, between 160 and 100~K for (001) and between 180 and 160~K for (111) films.
The sign change may be due to the strong magnetic fluctuations close to the N\'eel temperature,\cite{Cu-MNN_AHE} while further study is needed.
The AHC of (111) films is at most 21.5~($\Omega$~cm)$^{-1}$, which is almost one magnitude larger than that of (001) films.
By considering the AHC tensors of antiperovskite Mn$_3$GaN based on the $\Gamma_{4g}$ spin structure, the AHC for the (111) orientation is expected to be larger than for the (001) orientation.\cite{MGN_AHE}
Our results are qualitatively consistent with the theory, which thus, also supports that the obtained AHE is related to the intrinsic Berry phase.

\begin{figure}
\begin{center}
\includegraphics[width=\linewidth]{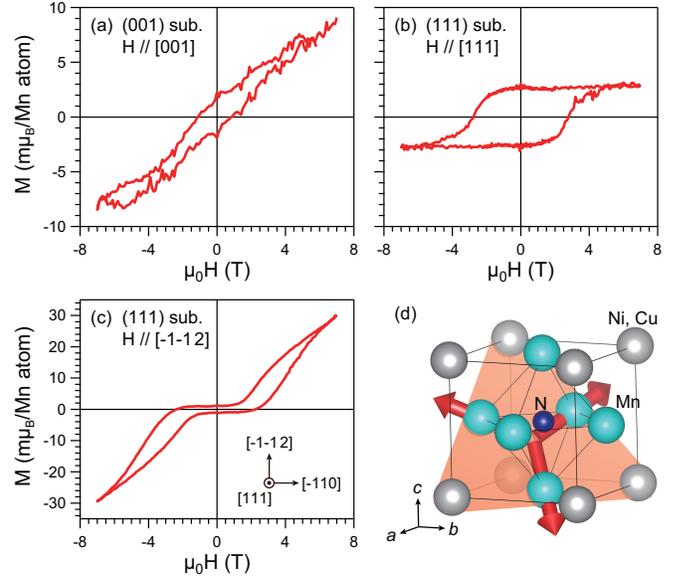}
\caption{
Magnetic properties of gas annealed Cu-MNN films at 50~K.
Out-of-plane hysteresis loops for Cu-MNN (a) (001) and (b) (111) films.
(c) In-plane hysteresis loop for Cu-MNN (111) films, for which the magnetic field is applied along the [-1-12] direction.
(d) Noncollinear AFM structure $\Gamma_{4g}$ in Cu-MNN, visualized using a software package.\cite{VESTA}
The gas annealing data for the (111) films are adapted from Ref.~\onlinecite{Cu-MNN_AHE}.
}
\label{fig:five}
\end{center}
\end{figure}

To discuss the different AHE loop shapes between (001) and (111) films, the magnetization as a function of magnetic field was measured.
\red{Figures~\ref{fig:five}(a) and \ref{fig:five}(b) show the out-of-plane magnetization hysteresis loop of Cu-MNN (001) and (111) films at 50~K, respectively.
Sharp flipping of the magnetic moment around a large coercive field ($H_c$) of about 3~T is found in the (111) films, while the (001) film is hard to saturate by a magnetic field with a small $H_c$ of about 1~T.}
The saturation magnetization was ${\sim}0.003\mu_{\rm B}$/Mn for (111) films, while no clear saturation was detected for (001) films.
An in-plane hysteresis loop for Cu-MNN (111) films at 50~K is shown in Fig.~\ref{fig:five}(c). 
\red{When the magnetic field was applied along [-1-12] direction, the obtained hysteresis loop was dramatically changed.
The obtained in-plane hysteresis loop is similar to that reported for a collinear AFM with a magnetic field parallel to the easy axis.}
\red{Therefore, the in-plane hysteresis loop implies that the Mn moments start to cant in (111) plane when the magnetic field is larger than threshold value of $2\sim3$~T.}
In contrast, since the magnetization due to symmetry-allowed spin canting in the $\Gamma_{4g}$ noncollinear state is along the [111] direction, \red{the Mn moments of Cu-MNN (001) films would not be sharp flipping when the magnetic field is applied along [001] direction}.
Besides, both the magnetization and the AHC of the (001) and (111) films have similar $H_c$, indicating that the AHE is directly related to the noncollinear AFM order.
From the AHE and magnetization results, Cu-MNN films have the $\Gamma_{4g}$ AFM order in the (111) plane, as shown in Fig.~\ref{fig:five}(d).

\begin{figure}[t]
\begin{center}
\includegraphics[width=\linewidth]{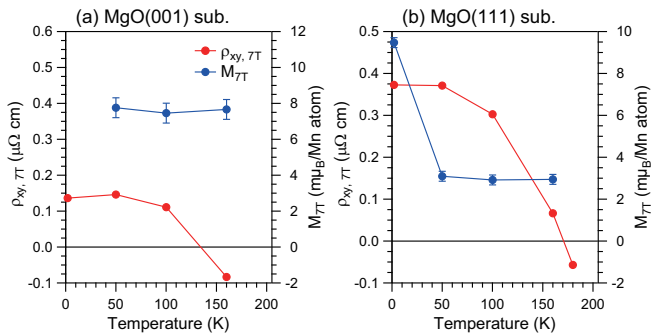}
\caption{
\red{Anomalous Hall resistivity and out-of-plane magnetization at 7~T as a function of temperature of gas annealed Cu-MNN (a) (001) and (b) (111) films.
For $\rho_{xy}$ of (001) films, the contribution of ordinarily Hall effect is subtracted. 
The gas annealing data for the (111) films are adapted from Ref.~\onlinecite{Cu-MNN_AHE}.}}
\label{fig:six}
\end{center}
\end{figure}

Finally, we would like to discuss the relationship between AHC and canted magnetization.
Figures~\ref{fig:six}(a) and~\ref{fig:six}(b) show the temperature dependence of the anomalous Hall resistivity and the out-of-plane magnetization of the (001) and (111) films, respectively. 
Though the canted magnetization does not change except 2~K, the anomalous Hall resistivity increases as decreasing temperature. 
In other words, there is no relationship between AHE and canted magnetization. 
On the other hand, the tilting angles from (111) plane estimated from saturation magnetization of (111) films are 0.07$^\circ$ at $50\sim160$~K and 0.20$^\circ$ at 2~K using $2.7\mu_{\rm B}$/Mn. 
According to the theoretical calculation of AHC in Mn$_3$Ir,~\cite{Hua_Chen_PRL} which have the same noncollinear AFM order as well as similar tilting angle, the AHC without the contribution from the noncollinear AFM order can be estimated to be less than 0.1~($\Omega$~cm)$^{-1}$ at the tilting angles below 0.2$^\circ$. 
In addition, we compared the AHC normalized by magnetization |$\sigma_{xy}$|/M with other Mn-based magnetic films in Table ~\ref{table:one}.
|$\sigma_{xy}$|/M of Cu-MNN (001) and (111) films exceeds 5000 and 300~($\Omega$$^{-1}$~cm$^{-1}$ $\mu_{\rm B}^{-1}$ Mn), respectively.
Those values are much larger than that of ferromagnetic Mn$_3$CuN(001) films.
In addition, we found that strained Mn$_3$NiN(001) films and polycrystalline Mn$_3$Sn films, where a large Berry curvature induces AHE, show similar values with Cu-MNN(001) and (111) films, respectively.
Therefore, although we cannot separate the contributions from canted magnetization and $\Gamma_{4g}$ noncollinear AFM order experimentally in this study, we conclude that most of AHE comes from $\Gamma_{4g}$ noncollinear AFM order rather than canted magnetization.

\begin{table}[b]
\begin{center}
\caption{
\red{Comparison of the normalized AHC by magnetization with similar Mn-based magnetic films.}
}
\begin{tabular}{m{12em}m{4em}c}
\hline \hline
 &  T  &	|$\sigma_{xy}$|/M\rule[0mm]{0mm}{4mm}\\
Film & (K) & ($\Omega$$^{-1}$~cm$^{-1}$ $\mu_{\rm B}^{-1}$ Mn) \\ \hline
Cu-MNN(111) & 50 & 5874 \rule[0mm]{0mm}{4mm}\\
Cu-MNN(001) & 50 & 339 \rule[0mm]{0mm}{4mm}\\
Mn$_3$CuN(001)~\cite{Mn3CuN_PRB} & 5 & 18\rule[0mm]{0mm}{4mm}\\
strained Mn$_3$NiN(001)~\cite{Mn3NiN_AHE} & 10 & 415\rule[0mm]{0mm}{4mm}\\
polycrystalline Mn$_3$Sn~\cite{Mn3Sn_poly1} & 300 & 3300\rule[0mm]{0mm}{4mm}\\
\hline \hline
\end{tabular} 
\label{table:one}
\end{center}
\end{table}


\section{CONCLUSIONS}
In conclusion, we successfully grew epitaxial Mn$_3$Ni$_{0.35}$Cu$_{0.65}$N films with the AHE due to nonzero Berry curvature.
We demonstrated the dependence of the magnetization and AHE on film orientation, and
found that the AHC of (111) films is an order of magnitude larger than that of (001) films.
Our films may have applications in AFM spintronics, such as spin-torque switching\cite{Yamane_SOT} and giant anomalous Nernst effect,\cite{Nernst} both of which have been predicted in the noncollinear AFMs.

\begin{acknowledgments}
The authors acknowledge Prof. Hua Chen for fruitful discussion.
The work in Nagoya was supported by the Japan Society for the Promotion of Science (KAKENHI grants 17K17801 and 17K19054) and the Hori Science and Arts Foundation.
The work in Augsburg was supported by the German Research Foundation through the priority program SPP 1666.
\end{acknowledgments}

\end{document}